# Supermode hybridization: a material-independent route towards record Schottky detection sensitivity using <0.05 μm$^3$ amorphous absorber volume

*Charles Lin, Pohan Chang, and Amr S. Helmy\**

The Edward S. Rogers Department of Electrical and Computer Engineering, University of Toronto, 10 King's College Road, Toronto, ON M5S 3G4, Canada

E-mail: a.helmy@utoronto.ca

**Abstract**

Schottky photodetectors are attractive for CMOS-compatible photonic integrated circuits, but the trade-off between photon absorption and hot carrier emission often compromises the detection fidelity and sensitivity. Here, we report a hybridization-based waveguiding effect that can improve the sensitivity of guided-wave Schottky detection by >200×. By hybridizing the supermodes guided by asymmetrical coupled plasmonic nanostructures, light-matter-interaction can be significantly enhanced even if the thickness of the light absorbing region is only a few nanometers, thus allowing the absorption and emission efficiencies to be simultaneously optimized for the first time. Using amorphous-based active junction with absorbing volume of only 0.031 μm$^3$, our detectors demonstrate record experimental sensitivity of -55 dBm, approaching that of crystalline-based Ge counterparts with >36x larger absorption region. The ability to maximize field overlap within small device volume and improve photodetection fidelity without using crystalline materials pave the way towards improved back-end-of-line interconnection, single photon detection, and 2D optoelectronics.

**MAIN TEXT**

The development of integrated photodetectors compatible with CMOS technology and Si photonics is critical towards optical sensing, computing, and high bandwidth density data transfer on chip (1). For these applications, detector sensitivity, reflected by the normalized photo-to-dark-current ratio, is one of the key factors that dictate the bandwidth and energy efficiency of the overall photonic integrated circuits (2). Although sensitive photodetection based on interband absorption of Si is widely employed in the visible range, it is not suitable for near-infrared detection due to Si's optical transparency above 1.1 μm. While defect-mediated mechanisms or nonlinear processes can also enable all-Si photodetection, their sensitivities are often compromised by significant leakage current and large input power requirement respectively (3-5). Currently, the heteroepitaxial growth of active Ge layer onto Si is deemed the most promising approach (6-9). However, these designs require complex doping, structuring, thermal, and cleaning treatments. Moreover, a thick crystalline Ge absorber layer (~1 μm) is required to achieve reasonable quantum efficiency. Since the epitaxy step can only be incorporated into the front-end-of-line CMOS processes, crystalline-based photonic devices will inevitably compete with on-chip electronics for valuable Si wafer real estate.

An emerging approach to detect infrared sub-bandgap optical radiation in Si is to employ the internal photoemission (IPE) process at Schottky interfaces (10-18). A representative Schottky detector and the associated IPE process are schematically illustrated in Fig. 1a and its inset respectively. The absorption of incident photons generates "hot" carriers, which have kinetic energies larger than those of thermal excitations at ambient temperatures. These excited hot carriers are transported towards the Schottky interface and may be emitted if the kinetic energy is greater than the barrier height ($\phi_B$). To-date, simple designs of IPE detectors based on overlaying metal films onto Si nanowires have demonstrated reduced RC product, shortened carrier transit time, and broadband (0.2-0.8 eV) operation. Using the deep-subwavelength field localization afforded by plasmonic structures, the electromagnetic field inside the metal absorber can be further enhanced for hot carrier excitation. Nonetheless, the quantum efficiency of Schottky detectors has thus-far remained low, as the emission probability is limited by the considerable momentum mismatch and poor overlap of the electron wavefunctions between the metal and Si (19).

Several strategies based on judicious junction design or absorber material optimization have been proposed to improve the quantum yield of guided-wave Schottky detectors. By roughening the Schottky interface, it is possible to minimize the impedance mismatch and produce higher photocurrents (20). Alternatively, the lowering of $\phi_B$ can also improve the quantum efficiency, but a corresponding increase in dark current ($I_d$) demands cryogenic operation (21). While resonant antenna structure can enable additional field-enhancement, this compromises the broadband nature of Schottky detection (22). Finally, by reducing the metal absorber thickness below the hot carrier mean free path, a better preservation of carrier energy and hence escape probability can be expected (Fig. 1b inset) (23). Nonetheless, the drawback is that the metal area within which electromagnetic energy interacts with electrons also become reduced, which compromises the speed and signal-fidelity of photoresponse. Such inability to simultaneously optimize the absorber region for both photon absorption and hot carrier emission thus prevents the practical implementation of compact and broadband Schottky photodetectors with sensitivity similar to their

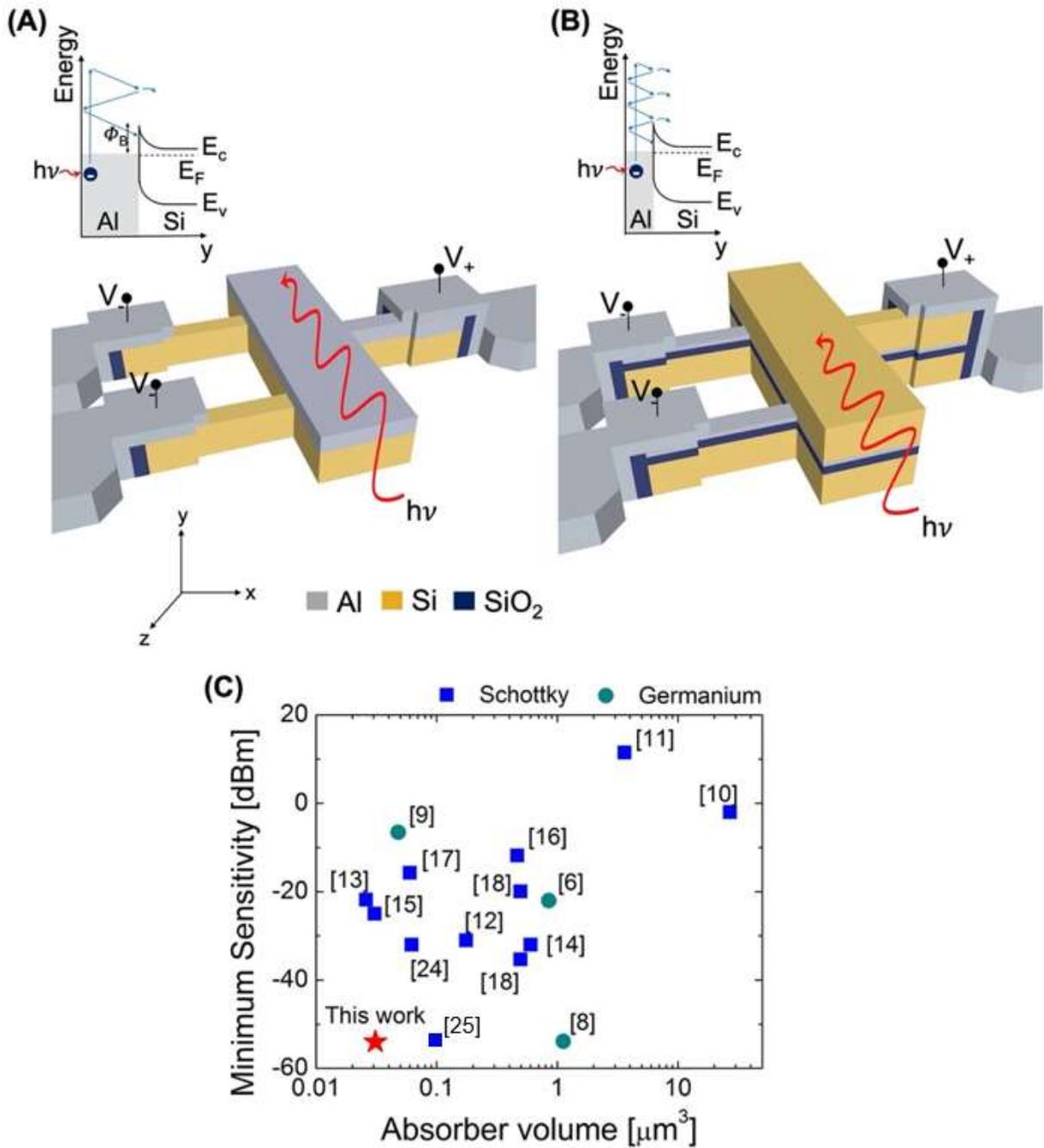

*Figure 1.* Schottky detector configuration and sensitivity. *(a)* Schematic of conventional guided-wave Schottky detector, consists of a thick metal layer that is overlaid onto a Si nanowire. Inset: energy band diagram of internal photoemission (IPE) at the Schottky interface. A photo-excited carrier may be emitted at the Schottky interface if it has energy greater than the Schottky potential

*($\Phi_B$) after being transported across the metal emitter layer.* **(b)** *Schematic of our Schottky detector, which is a multi-layer structure that supports supermodes that have been hybridized. Inset: energy band diagram of IPE within a thin metal emitter (23). Hot carriers may reflect off the internal metal surface multiple times before being emitted.* **(c)** *Comparison of experimentally demonstrated guided-wave Schottky and Ge detectors at λ = 1.55 μm.*

Ge counterparts (Fig. 1c). Moreover, the approach of optimizing the optoelectronic properties of the active Schottky region can limit material selection and demand costly front-end-of-line CMOS integration.

Recently, we demonstrated an integrated Schottky detector with record -54 dBm sensitivity using a 10 nm thick metal absorber (Fig. 1b) (24, 25). With a single active junction formed with amorphous materials, the design is backend compatible and can couple to Si nanowires with only 1.5 dB loss, thereby demonstrating the possibility of plasmonic–photonic integration on top of Si chips without costly process modifications. Furthermore, the detector can potentially enable FPGA-like, reconfigurable optical circuits, as the same waveguide architecture can be utilized for filtering and modulation by changing the bias configuration (26, 27). Nonetheless, the physical mechanism that the photodetector utilized to overcome the absorber dilemma has yet been reported.

In this work, we report a waveguiding effect that can enable non-resonant, ultra-sensitive Schottky detection using a <0.05 μm$^3$ absorber volume. By introducing structural asymmetry into coupled-mode plasmonic waveguides, it will be shown that an induced hybridization between the guided supermodes can improve the efficiency of metal absorbers by several orders of magnitudes. This hybridization-based enhancement does not require any optimization of the optoelectronic properties of the active junction and can be utilized even if the light absorbing region is purely amorphous-based and has a thickness of only a few nanometers. The amorphous devices reported here demonstrate a record experimental sensitivity down to -55 dBm, approaching that of crystalline-based Ge counterparts that rely on strong electron-hole pair generation and have >36× larger absorption region.

Before demonstrating the possibility of maintaining small Schottky active region while increasing the achievable optical absorption, it is necessary to elucidate the importance of minimizing the size of the active absorbing region. Figure 2 shows the properties of a CMOS-compatible Al-Si Schottky junction as a function of its metal emitter/absorber thickness ($t$). The details of the detection model can be found in *Materials and Methods*. It follows the work by Scales et al. (23) but has been expanded to account for carrier thermalization. Although it provides a first-order estimate of photodetector efficiency since the effect of carrier momentum mismatch has not been included, its predictions are comparable with previous experimental results and therefore sufficient for illustrating the performance benefit of our proposed structure (9,10). Specifically, it is observed that the probability of a hot electron being excited by surface plasmon polaritons (SPP) and then transported to the Schottky interface ($P_{generation}$) improves with increasing $t$ (Fig. 2a). However, the probability of hot electron emission ($P_{emission}$) improves with decreasing $t$. This is because a reduced transport distance can allow hot electrons to reflect off of the internal metal surfaces and traverse multiple roundtrips in the metal and still reach the interface with sufficient kinetic energy to be emitted (23). From Fig. 2b, it is observed that this enhancement from thin-film emitters requires $t$ to be <20 nm and is optimized at $t = 2.5$ nm, where the internal quantum efficiency (IQE) is calculated to be 7.5%. The waveguide analysis and experimental demonstration reported in this work will mainly focus on photodetectors with $t = 10$ nm (IQE ~5%), as it is minimum reproducible layer thickness achievable using backend-compatible

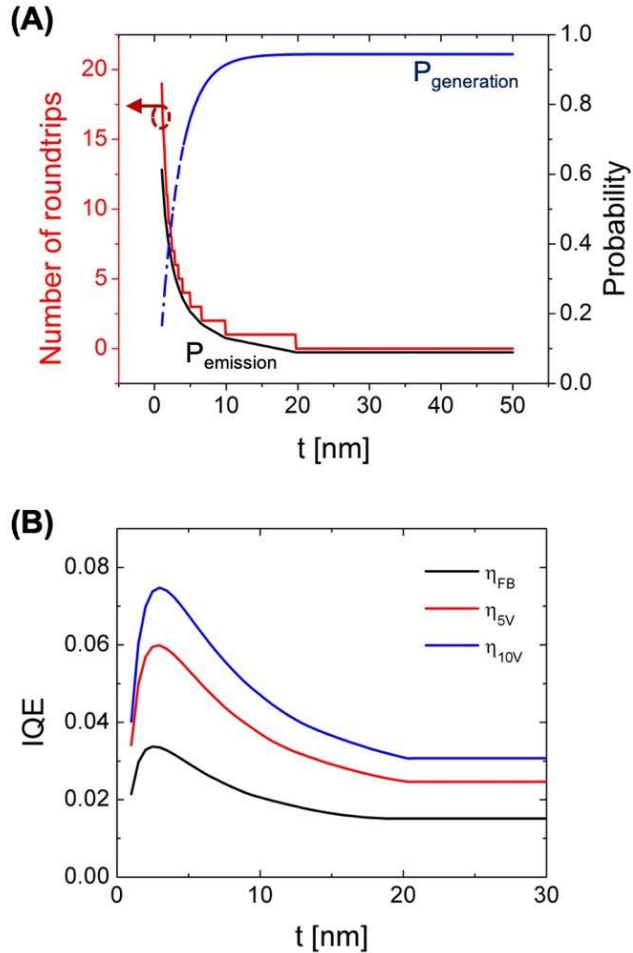

*Figure 2.* Modelling of internal photoemission at Al-Si Schottky junction. *(a)* Probability of hot electrons being excited and transported to the Al-Si interface ($P_{generation}$) and the probability of hot electron emission ($P_{emission}$) at the interface. Number of round trips that a hot carrier can reflect off of the internal metal surfaces before emission is also plotted. Details of the internal photoemission modelling can be found in the Materials and Methods section. The $\Phi_B$ for the Al–Si junctions are assumed to be 0.54 eV and image-lowering, carrier attenuation, and carrier reflection at the internal metal interface are considered. The effect of electron wavefunction mismatch in the metal and Si is not included. Note that the experimental material parameters will *likely deviate from the ones used in the modelling.* ***(b)*** *Internal quantum efficiency of hot electrons as a function of Al emitter/absorber thickness (t) for flat band voltage, 5V, and 10V.*

sputtering or evaporation processes.

A photodetector waveguide structure that can enhance the sensitivity of guided-wave Schottky detector is schematically shown in Fig. 3a. It is a multilayer stack constructed using CMOS-compatible materials that can be deposited via back-end-of-line processes. Electrically, it consists of the Si-Al Schottky photodetection junction that is joined to a bottom Al-SiO$_2$-Si metal-oxide-semiconductor (MOS) structure through a common Al layer. Optically, the Schottky interface supports surface plasmon polariton (SPP) modes while the MOS region supports hybrid plasmonic waveguide (HPW) modes (27). If *t* is less than the skin depth, the mutual perturbation between the evanescent fields of the SPP and HPW modes can lead to the formation of transverse-magnetic antisymmetric ($TM_{S,i}$) and symmetric ($TM_{L,j}$) supermodes, where indices *i* and *j* represent the mode order (Fig. 3b). Similar to that of coupled oscillator systems, the former are highly dissipative hard modes that correspond to constructive interference and hence have increased field overlap at the metal region while the latter are low-loss soft modes that relate to destructive interference and therefore reduced field-metal overlap (28). Although the $TM_{S,i}$ supermodes can provide modal absorption up to several dB/μm and thus highly suitable for photodetection, the lack of efficient excitation mechanisms prevents their practical utilization. On the contrary, the $TM_{S,i}$ supermodes can be efficiently excited with Si nanowires and fibers, but are generally only used for components such as modulators or resonators that require minimal insertion loss (26,27).

The absorption efficiency of the Schottky metal emitter can be maximized through the hybridization between the $TM_S$ and $TM_L$ supermodes. In order to achieve hybridization, two

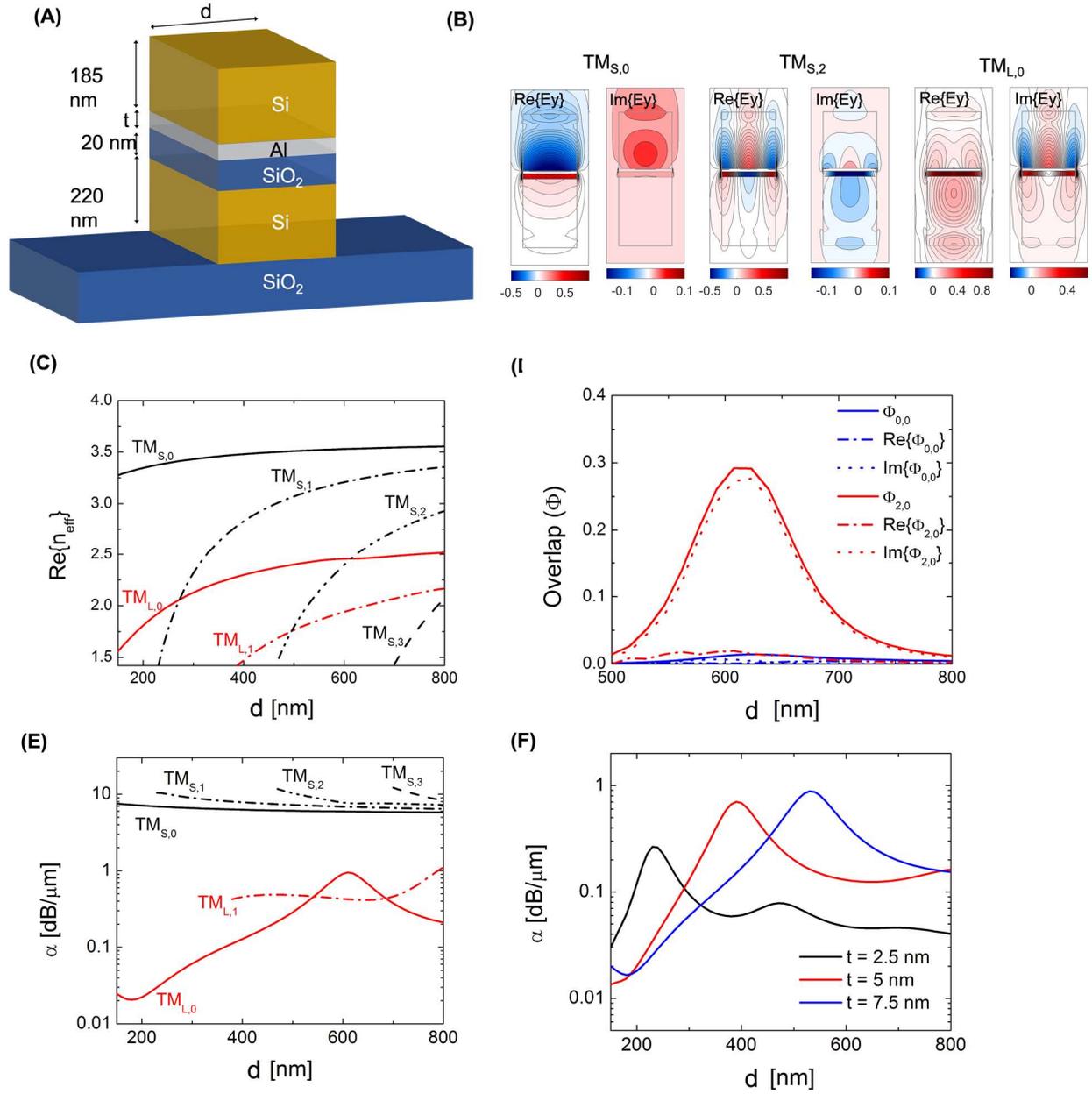

*Figure 3. Supermode hybridization within multi-moded, asymmetrical coupled-mode structure. (a) Schematic of the plasmonic photodetector waveguide. (b) Field distributions of the selected supermodes at λ = 1.55 μm and waveguide width (d) of 620 nm (only the dominant $E_y$ field component is plotted). The top Si layer thickness is fixed at 185 nm. (c) Effective indices, (d) field overlap strength, and (e) absorption coefficient (α) of the guided supermodes as a function of*

*waveguide width (d). **(f)** α of the $TM_{S,0}$ supermode for different metal layer thicknesses (t). The top Si layer thickness is 190, 188, and 186 nm for t = 2.5, 5, and 7.5 nm respectively.*

eigenmodes need to be phase matched and have sufficient spatial field overlap. Based on numerical dispersion calculations shown in Fig. 3c, it is observed that phase matching condition cannot be established between the fundamental $TM_{S,0}$ and $TM_{L,0}$ supermodes. On the other hand, the overlap strength can be calculated by (29,30):

$$\Phi_{i,j} = \frac{|\iint \vec{E}_{S,i} \vec{E}_{L,j}^* d\vec{S}|^2}{\iint |\vec{E}_{S,i}|^2 d\vec{S} \iint |\vec{E}_{L,j}|^2 d\vec{S}} = \frac{|\iint (\vec{E}_{S,i}' \vec{E}_{L,j}^{*'} + \vec{E}_{S,i}'' \vec{E}_{L,j}^{*''})^2 d\vec{S}|^2}{\iint |\vec{E}_{S,i}|^2 d\vec{S} \iint |\vec{E}_{L,j}|^2 d\vec{S}} + i \frac{|\iint (\vec{E}_{S,i}' \vec{E}_{L,j}^{*''} - \vec{E}_{S,i}'' \vec{E}_{L,j}^{*'})^2 d\vec{S}|^2}{\iint |\vec{E}_{S,i}|^2 d\vec{S} \iint |\vec{E}_{L,j}|^2 d\vec{S}},$$

where $E_{S,i}$ and $E_{L,j}$ are the real and imaginary field distributions of supermodes respectively. Moreover, $\Phi_{i,j}$ can be separated into real and imaginary parts, which are known as coherent and dissipative coupling strength respectively (30). Figure 3d reveals that since the $TM_{S,0}$ and $TM_{L,0}$ supermodes can only overlap coherently up to ~1.5%, the field-matching condition also cannot be met. This is because the $TM_{S,0}$ and $TM_{L,0}$ supermodes are both dominated by their $Re\{E_y\}$ field component, which have antisymmetric and symmetric field distributions with respect to the metal respectively (Fig. 3b). However, by increasing the width of the waveguide structure (*d*) such that it becomes multi-moded, it becomes possible for supermode hybridization to take place, albeit between supermodes of different mode order. Specifically, the $TM_{S,2}$ and $TM_{L,0}$ supermodes can be phase-matched with $\Phi_{2,0}$ as high as 30% at *d* = 620 nm (Fig. 3d). As shown in the evolution of the $TM_{L,0}$ supermode profile in *Suppl. Mat. 1*, the magnitude of the $Im\{E_y\}$ field component increases with *d*. Concurrently, the field distribution on top of the Schottky interface also evolves from single lobe distribution into a three-lobe distribution, which matches well with the $Re\{E_y\}$ field component of the $TM_{S,2}$ supermode (Fig. 3b). As a result, the two supermodes starts to couple dissipatively and an anti-crossing behavior is found between their absorption curves near $d = 620$ nm, which is characteristic of mode hybridization (28). As a result, without relying on optimizing the active material junction and using a 10 nm metal emitter, an otherwise low-loss $TM_{L,0}$ supermode can be hybridized to provide strong absorption of 1 dB/µm (Fig. 3e).

To the author's knowledge, this is the first report of hybridization between supermodes guided within a single waveguide element instead of between neighboring waveguide structures. These hybridized modes are engineered to exhibit the strong optical absorption only attainable in $TM_S$ supermodes but still maintain the field distribution typical of a $TM_L$ supermode, one that can interface efficiently with Si photonics. As shown in Suppl. Mat. 2, enhancement mechanism is non-resonant and athermal, as strong absorption can be maintained from $\lambda = 1.3$ to 1.6 µm. Moreover, orders of magnitude absorption enhancement can be obtained even if $t$ is reduced to a few nanometers (Fig. 3f). Specifically, modal absorption up to 0.3 dB/µm is possible for $t = 2.5$ nm. Note that hybridization can be induced between any combinations of $TM_{L,j}$ and $TM_{S,j+k}$ supermodes, where $k$ is a positive even integer. For example, an absorption peak is also observed by hybridizing the $TM_{L,1}$ supermode with the $TM_{S,3}$ supermode at $d = 800$ nm (Fig. 3e).

The magnitude of the enhanced absorption efficiency is controlled by the amount of structural asymmetry inherent to the coupled-mode structure. Figure 4 displays how the $TM_{L,0}$ modal absorption evolves as a function of $w$ for photodetector waveguide configurations with increasing level of asymmetry. For a coupled-mode structure consisted of two identical Schottky junctions and surrounded by homogenous $SiO_2$ superstrate and substrate, modal absorption does not vary significantly as the waveguide is widened and approaches an asymptote of 0.0378 dB/µm. However, by simply switching the superstrate from $SiO_2$ to air, a gradual absorption peak starts to manifest. Next, by introducing a 20 nm $SiO_2$ waveguide layer such that the field distributions above and below the metal emitter become asymmetrical (Fig. 3b), an abrupt change in absorption is obtained, which can increase further by increasing the thickness of the $SiO_2$ layer. This trend is analogous to the hybridization between the $TM_0$ and $TE_1$ modes of a Si nanowire that occurs with increasing sidewall angle or refractive index contrast, though the modes couple coherently instead of dissipatively (31,32). It is important to highlight that the construction of photodetector waveguide using a merger of two dissimilar plasmonic structures provides many degrees of freedom for optimization. For example, by finetuning the thickness of the top Si layer in addition to modifying $d$, $TM_{L,0}$ absorption can be further improved to 2.04 dB/μm (*Suppl. Mat. 3*).

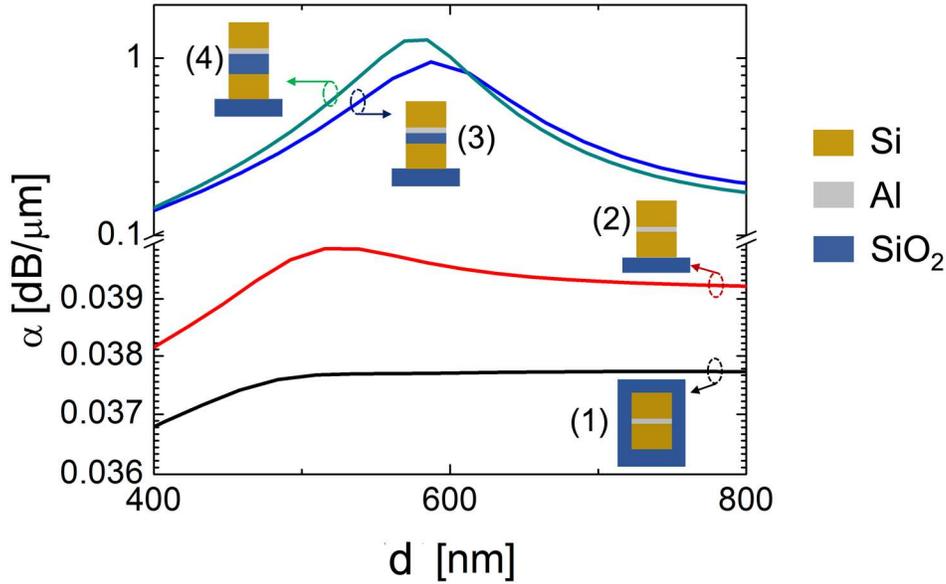

*Figure 4.* Effect of structural asymmetry on supermode hybridization. Absorption coefficient (α) of the $TM_{L,0}$ supermodes for four different coupled-mode waveguide configurations with increasing level of asymmetry: (1) Completely symmetrical structure ($SiO_2$ superstrate/220 nm Si/10 nm Al/220 nm Si/$SiO_2$ substrate). (2) Asymmetrical structure with air superstrate (air/230 nm Si/10 nm Al/220 nm Si/$SiO_2$ substrate). (3) Asymmetrical structure with an additional $SiO_2$ layer (air/185 nm Si/10 nm Al/20 nm $SiO_2$/220 nm Si/$SiO_2$ substrate). (4) Asymmetrical structure with thicker $SiO_2$ layer (air/175 nm Si/10 nm Al/40 nm $SiO_2$/220 nm Si/$SiO_2$ substrate).

To demonstrate the capability of supermode hybridization, Schottky photodetectors with 10 nm emitter thickness have been fabricated and tested (Fig. 5a). The bottom Si layer of the fabricated device is crystalline since standard Si-on-insulator wafers are utilized for proof-of-concept. However, the active junctions are implemented using Al and undoped α-Si sputtered at room temperature. To minimize disturbance to the optical mode, 200 nm thick Al contacts are made to the α-Si and Al waveguide layers through finger structures. The Al emitter is biased at a negative potential with respect to the Al collector on top. The photodetectors are excited by first coupling light from a 750 nm Si nanowire into a 200 nm wide detector section, which then tapers over 500 nm to the optimal width of 620 nm. Using the cut-back method, the coupling efficiency from an input Si nanowire into the photodetector is measured to be 75%.

The current-voltage characteristics of a 5 μm long photodetector is shown in Fig. 5b. The $I_d$ curve follows the trends expected from thermionic emission theory (33), with reach-through voltage and flat-band voltage of 2.1 V and 3 V respectively. Owing to compact contact area of only 1μm$^2$, $I_d$ only reaches ~1 nA to bias up to 10 V. Once optical illumination at $\lambda$ = 1.55 μm is applied, photocurrent ($I_p$) of 20 nA is measured and a linear relationship with input optical power is observed (Fig. 5c), suggesting that measured response does not originate from two-photon absorption (34). With a maximum electric field strength of ~50V/μm across the α-Si layer, the electron ionization coefficient and avalanche multiplication factor can be calculated to be $10^{-3}$ μm$^{-1}$ and 1.22 respectively (18). Thus, avalanche multiplication is not expected to contribute significantly to the photoresponse.

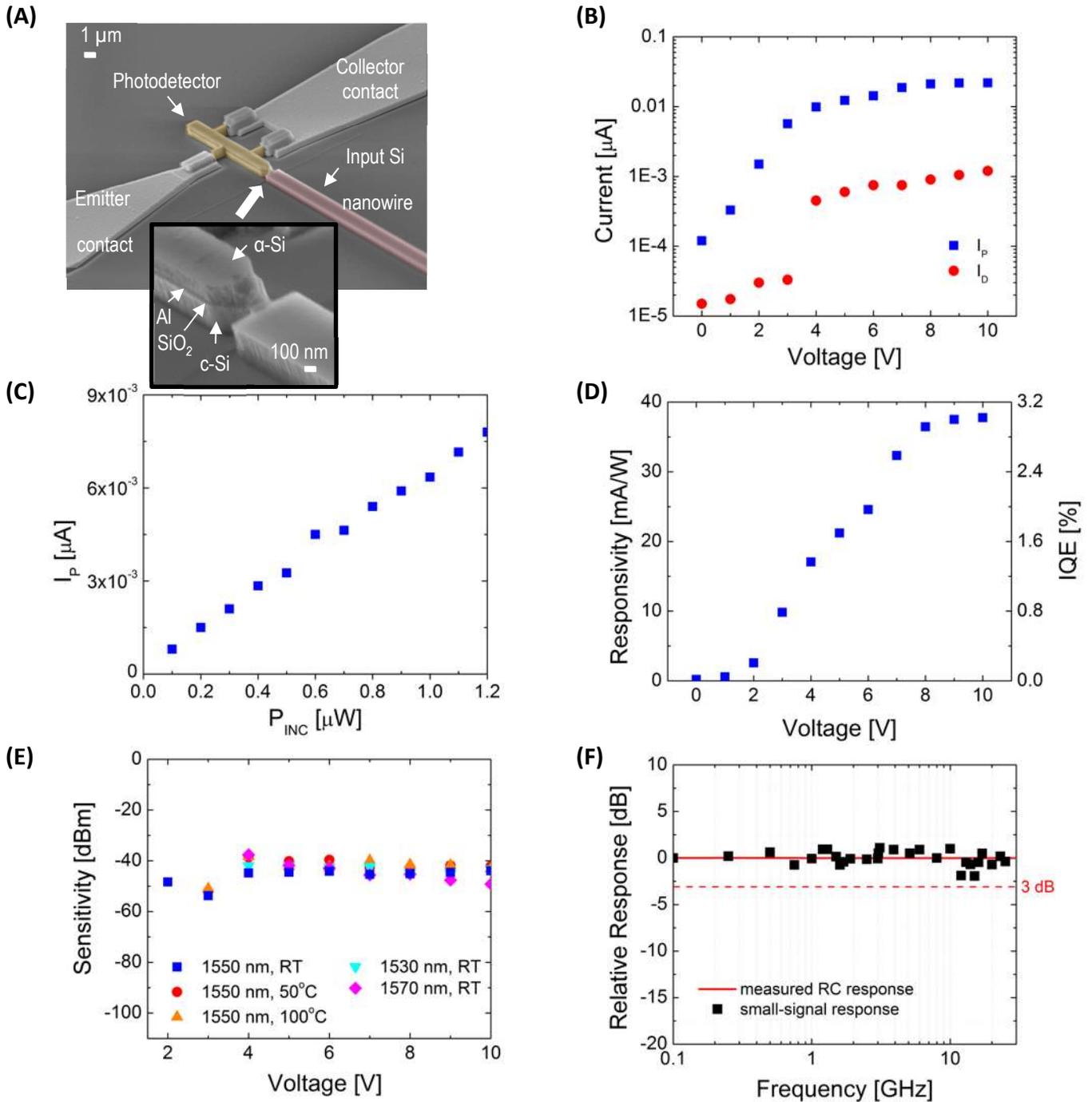

*Figure 5. Experimental photodetector results. (a) False-colored SEM image of a 5 μm photodetector. (b) Detector dark current ($I_d$) and photocurrent ($I_p$) at λ=1.55 μm. (c) $I_p$ as a function of input optical power. (d) Responsivity and internal quantum efficiency at room temperature and λ=1.55 μm. (e) Sensitivity of the photodetector at different operating wavelengths*

*and temperature conditions. The devices are tested up to equipment limitation of 1.57 μm but is expected to be operational up to 1.8 μm based on simulation. **(f)** Frequency response measured up to 26 GHz and fitted to its RC response, modeled using a 50 Ω load resistance and 12 fF measured capacitance. The measurement is normalized to the response at 100 MHz.*

Harnessing the strong absorption afforded by supermode hybridization, previously unattainable photodetection efficiency and sensitivity have been demonstrated (Fig. 5d). The maximum responsivity and IQE are measured to 38 mA/W and 3.2 % respectively. More importantly, a record Schottky detection sensitivity of –55 dBm is measured at 3V bias (Fig. 5e). This is comparable to that of crystalline-based Ge detectors with 1 μm$^3$ absorber volume but achieved using amorphous active junction and an absorber volume of only 0.031 μm$^3$ (Fig. 1c). Furthermore, the detector exhibits athermal behavior as strong sensitivity can be maintained across broad bias conditions (2-10V), wavelength range (1.53-1.57μm), as well as substrate temperatures (20-100ºC). As wavelength shortens, the decreased optical power coupled into the device is compensated by higher IQE. On the other hand, while thermionic emission dictates that an elevated temperature will lead to an increase in $I_d$, sensitivity remain unchanged as a result of a corresponding increase in $I_P$ due to temperature-induced lowering of $\Phi_B$ in α-Si (35).

Finally, using available instrumentation, the speed of the photodetector is confirmed up to 26 GHz (Fig. 5f). With measured parasitic capacitance of 12 fF and assuming 50Ω resistance, the RC response of the fabricated photodetectors have a calculated RC cut-off frequency of 265 GHz. Although the detector speed will also be restricted by carrier transit time, the limitation can be alleviated in future device generations through controlled doping profiles as well as reducing the α-Si layer thickness.

As shown in the detailed comparison of experimental guided-wave Schottky detectors in *Suppl. Mat. Table S1*, inducing supermode hybridization with asymmetrical coupled-mode structure is the most effective method towards maximizing detector performance. Without the need to optimize the material properties of the active junction, this waveguiding approach allows the enhanced emission probability associated with thin-film metal emitters to be fully utilized without sacrificing absorber efficiency. This in turn enables athermal Schottky photodetectors to be implemented with compact footprint (5 μm) as well as minimal idle current (<1 nA), ultimately leading to record sensitivity (-55dBm), responsivity per volume (871 mA/Wμm$^3$), and RC bandwidth (265 GHz). In comparison, designs based on thin layers of silicide emitters demonstrated weaker sensitivity (–30 dBm) while detectors that incorporated graphene injection layer demonstrated responsivity per volume of 748 mA/Wμm$^3$. With improved waveguide fabrication, the detector also enables 1 dBm improvement in sensitivity but is 3x shorter compared to our prior work (25).

Although the use of amorphous materials is associated with smaller carrier mobility, lower saturation velocity, as well as higher recombination (9), our integrated Schottky detectors have sensitivity that is competitive compared that of crystalline-based Ge detectors for the first time, achieved using an active volume of 0.031 μm$^3$ that is 36 times smaller. Concurrently, the reduced dark current level point towards better photodetection fidelity as well as smaller idle power consumption and larger bandwidth. Overall, these detectors are highly suitable for low-cost, CMOS back-end-of-line processes and optoelectronic integration, away from expensive transistor levels and avoid modifications of front-end-of-line CMOS manufacturing.

Although supermode hybridization is used for enhancing the absorption of Schottky emitters in this work, it is a versatile effect that can complement other mechanisms or be utilized for other applications. For example, further improvement in detector efficiency is expected if the coupled-mode structure becomes embedded in high doped semiconductor with strong free carrier absorption, incorporates an additional graphene injection layer, or operates in the avalanche gain regime (18, 36, 37). The high-fidelity response is not only beneficial for interconnection, but also critical for near- and mid-infrared imaging, photovoltaic solar energy conversion, and (bio)chemical sensing (38). The ability of our structure to maximize field overlap within a small device volume can also be leveraged towards 2D material-based optoelectronics or single photon detection using superconducting nanowire, as these functionalities are currently limited by the size of the device region (18,39,40).

**METHODS**

Photodetector electrical modeling

The photodetection model presented here follows the formulism reported in references (23) and (41). Specifically, the probability of a hot electron to be excited within a metal layer is proportional to the intensity of the surface plasmon mode and given by:

$$P_1 = e^{\frac{-x}{\delta}},$$

where x is the spatial coordinate along the metal layer and δ is the skin depth the SPP mode in the metal region (~25 nm for Al at λ=1.55 μm). Once generated, hot electron may undergo inelastic collisions, which reduces the probability of reaching the Schottky interface as described by:

$$P_2 = e^{\frac{-x}{L}},$$

where L is the mean free path (~100 nm for Al at λ=1.55 μm). Thus, the total probability of a hot electron being generated and reach the Schottky interface with a metal emitter of thickness t is

$$P_{generation} = \frac{1}{\delta}\int_0^t P_1 P_2 dx.$$

For a thin-film detector with a single Schottky interface, the internal quantum efficiency is given by:

$$\eta_i = \frac{1}{h\upsilon} P_{generation} \int_{\Phi_B}^{h\upsilon} P^t(E_0) dE_0,$$

where $P^t(E_0)$ is the sum of carrier emission probability with energy level $E_0 > \Phi_B$, given as:

$$P^t(E_0) = P_0 + (1-P_0)P_1 + (1-P_0)(1-P_1)P_2 + \cdots P_n \prod_{k=0}^{n-1}(1-P_k).$$

The probability $P_k$ for energy level $E_k$ after k round trips between the two interfaces of the metal is:

$$P_k = \frac{1}{2}\left(1 - \sqrt{\frac{\Phi_B}{E_k}}\right)$$

$$E_k = E_0 e^{-2k\frac{t}{L}}.$$

Finally, the total number of round trips between the interfaces before $E_k$ falls below $\Phi_B$ is given by:

$$n = \frac{L}{2t} \ln\left(\frac{E_0}{\Phi_B}\right).$$

Photodetector optical simulation

The effective index and propagation loss of the waveguide mode were calculated via 2D finite element method simulation using the commercial Lumerical Mode Solutions software. Metallic boundary conditions were used to terminate the 2 μm by 2 μm computational domain. Grid sizes of 0.1, 1, and 2.5 nm were used to mesh the Al thin film, the SiO$_2$ spacer layer, and the rest of the waveguide structure, respectively. The top Si layer was taken to be crystalline in the simulations. The refractive indices of the materials at λ = 1.55 μm are as follows: $n_{Si}$ = 3.4784, $n_{SiO2}$ = 1.44, and $n_{Al}$ = 1.44 + 16i.

Photodetector fabrication

The photodetectors were fabricated on SOITEC wafer with 220 nm thick crystalline Si layer. The 185 nm thick top Si layers was deposited using AJA International ATC Orion 8 Sputter Deposition System in Ar plasma. An undoped Si target was used with bulk resistivity > 1Ω−cm. The 20 nm thick $SiO_2$ and 10 nm thick Al were deposited using Angstrom Nexdep Electron Beam Evaporator. The root-mean-square roughness of the deposited stack was measured using atomic force microscopy to be 1.49 nm. Multiple lithography steps were required to define the photodetectors and they were carried out using the Vistec EBPG 5000+ Electron Beam Lithography System. For device patterning, the Si and $SiO_2$ layers were etched using the Oxford Instruments PlasmaPro Estrelas 100 DRIE System with a $SF_6$:$C_4F_8$ mixed-gas. Electrical contacts were fabricated by sputtering 200 nm thick Al onto the devices. The experimental device dimensions were confirmed with scanning electron microscopy.

Photodetector characterization

The set-up for measuring the photodetector's DC response was as follows: a continuous-wave from JDS Uniphase SWS15101 tunable laser was chopped at 377Hz and coupled into the sample. At the same time, a DC bias was applied across the photodetector using a Keithley 26013 source meter. The current generated by the photodetector was collected using a Cascade Infinity probe, converted into photo-voltage using a 10 kΩ resistor, and then measured using a Stanford Research System SR830 DSP Lock-In Amplifier.

AC measurement were taken using the NI vector signal generator (VSG) and vector signal analyzer (VSA). On the input side, the VSG stimulus was amplified via Erbium-doped fiber amplifier and drove a Fujitsu LiNbO$_3$ Mach-Zehnder modulator, generating intensity-modulated optical signals. On the output side, the photocurrent was collected with a Cascade Infinity probe and amplified via AMF-4F-060180 amplifier before feeding into the VSA.

**AUTHOR CONTRIBUTIONS**

A.S.H., C.L. and P.H.C. have designed the structures. C.L. and P.H.C. have electromagnetically modeled the structure. C.L. has fabricated the samples. C.L. and P.H.C. have characterized the devices. A.S.H., C.L. and P.H.C. have analyzed the measurements and wrote the manuscript.

**NOTES**

The authors declare no competing interests

**SUPPORTING MATERIALS AVAILABLE**

The following files are free of charge.

- Field profile, wavelength-dependence, and 2D optimization of the photodetector waveguide.
- Comparison of experimental Schottky detectors.